\documentclass[twocolumn,showpacs,prepringnumbers,amsmath,amssymb,prb]{revtex4}
\usepackage{graphicx}% Include figure files
\usepackage{dcolumn}% Align table columns on decimal point
\usepackage{bm}% mathbf math
\usepackage{subfigure}
\usepackage{wrapfig}
\begin{document}
\preprint{APS/123-QED}
\title{Electrothermal flow in Dielectrophoresis of Single-Walled Carbon Nanotubes}% Force line breaks with \\

\author{Yuan Lin$^1$}

\affiliation{1: Department of Mechanics, Royal Institute of
Technology, Osquare Backe 18, Stockholm, 100 44, Sweden\\
2: Department of Mechanical Engineering, The University of Tokyo,
7-3-1 Hongo, Bunkyo-ku, Tokyo, 113-8656, Japan}

\author{Junichiro Shiomi$^2$}
\affiliation{1: Department of Mechanics, Royal Institute of
Technology, Osquare Backe 18, Stockholm, 100 44, Sweden\\
2: Department of Mechanical Engineering, The University of Tokyo,
7-3-1 Hongo, Bunkyo-ku, Tokyo, 113-8656, Japan}

\author{Shigeo Maruyama$^2$}
\affiliation{1: Department of Mechanics, Royal Institute of
Technology, Osquare Backe 18, Stockholm, 100 44, Sweden\\
2: Department of Mechanical Engineering, The University of Tokyo,
7-3-1 Hongo, Bunkyo-ku, Tokyo, 113-8656, Japan}

\author{Gustav Amberg$^{1,}$}
\email{gustava@mech.kth.se}
\affiliation{1: Department of Mechanics, Royal Institute of
Technology, Osquare Backe 18, Stockholm, 100 44, Sweden\\
2: Department of Mechanical Engineering, The University of Tokyo,
7-3-1 Hongo, Bunkyo-ku, Tokyo, 113-8656, Japan}


\begin{abstract}
We theoretically investigate the impact of the electrothermal flow on the dielectrophoretic separation of single-walled carbon nanotubes (SWNT). The electrothermal flow is observed to control the motions of semiconducting SWNTs in a sizeable domain near the electrodes under typical experimental conditions, therefore helping the dielectrophoretic force to attract semiconducting SWNTs in a broader range. Moreover, with the increase of the surfactant concentration, the electrothermal flow effect is enhanced, and with the change of frequency, the pattern of the electrothermal flow changes.
%It is shown that under some typical experimental conditions of separation of SWNTs by dielectrophoresis, the electrothermal flow is a dominating factor in determining the motion of SWNTs.
It is shown that under some typical experimental conditions of dielectrophoresis separation of SWNTs, the electrothermal flow is  a dominating factor in determining the motion of SWNTs.
\end{abstract}

%Use showkeys class option if keyword
\maketitle
\section{Introduction}
Single-walled carbon nanotubes are key materials in nanotechnology as potential candidates for diverse applications owning to their extraordinary mechanical, thermal, optical and electrical properties \cite{Saito}. On exploring the utility of the electrical properties, one of the current critical challenges is the separation of metallic (m-SWNTs) and semiconducting SWNTs (s-SWNTs). Among various post-synthesis separation methods devised and applied \cite{krupke2003,chen,Chattopadhyay,Zheng,Maeda}, dielectrophoresis separation has been demonstrated to be possible with high selectivity and simplicity
\citep{krupke2004}. Furthermore, DEP also allows us to deposit SWNTs to selected sites to construct electric devices \cite{zhibin1}. While the complexity in the optical measurements gives rise to difficulties in interpretation of the obtained spectra \cite{Baik}, deeper understanding in the transport dynamics of SWNTs under DEP operation is of a great importance. It also allows us to discuss the efficiency and possibility for optimization of system designs and working conditions. Here, numerical simulations should serve as a powerful tool since the dynamical observation of SWNTs transported in the suspension is extremely challenging in experiments.

On modeling a SWNT as a prolate ellipsoid, the DEP force on the SWNT in an inhomogeneous AC electric field $\bf E$ is calculated as \cite{krupke2004},
\begin{equation}
\mathbf F_{\text{DEP}}=\frac{\pi
ab^2\epsilon_\text{m}}{12}\alpha\nabla|\mathbf E|^2, \label{eqn1}
\end{equation}
where $\alpha=\displaystyle Re[\frac{\epsilon_p^*-\epsilon_\text{m}^*}{\epsilon_\text{m}^*+(\epsilon_p^*-\epsilon_\text{m}^*)L_p}]$, $\quad\epsilon_{m,p}^*=\displaystyle\epsilon_{m,p}-i\frac{\sigma_{\text{m,p}}}{\omega}$
, $\omega$ is the frequency of the AC field and $\epsilon_{\text{m,p}}^*$ is the complex dielectric permittivity. Here the subscripts $m$ and $p$ denote the medium and particle. The constants $a$ and $b$ denote half the tube-length and tube-radius. The depolarization factor
$L_p=\displaystyle\frac{b^2}{2a^2e^3}[ln(\frac{1+e}{1-e})-2e]$,
$e=\displaystyle\sqrt{1-\frac{b^2}{a^2}}$.

On calculating the DEP force of SWNTs using parameters from electrical measurements of suspended pure SWNTs \cite{Dai}, the force on m-SWNTs becomes several orders higher than that on s-SWNTs, indicating that the separation should be easy. Furthermore, numerical calculations suggested that DEP forces of both m-SWNTs and s-SWNTs are insensitive to variations in frequency of the AC electric field and surfactant concentration within the accessible range in experiments \cite{dimaki}. This does not agree with the experimental results where the success of DEP separation strongly depended on the frequency and the surfactant concentration \cite{krupke2004}. Experiments showed that the separation was successful only above some certain frequencies below which both m-SWNTs and s-SWNTs were found on the electrodes. Although the complete picture of the transport dynamics is yet to be revealed, Krupke {\it et al} \cite{krupke2004} attributed the discrepancy to the electrical conduction through the surrounding surfactant. They have represented the surface conduction effect with the effective electrical conductivity which was obtained by fitting the solution of Eq.~(1) to the experimental results.

Previous numerical simulations on DEP separation of SWNTs have usually considered only the DEP forces and the Brownian motion assuming that the bulk flow velocity is zero.
Although the basic concept of DEP-separation in principle is simple, the system involves effects which may cause bulk flow motions such as electroosmosis, thermal convection, electrothermal flow.
The electroosmosis effect is only important at low frequencies (less than $10^4$ Hz, which is well
below the frequency applied in experiments) \cite{morgan}, and the thermal convection is expected to be negligible in micro and nanoscales. On the other hand, the electrothermal effect is known to be  substantial in micro systems \cite{morgan,ramos}. Electrothermal flow is driven by a body force caused by electric field acting on gradients in permittivity and/or conductivity due to a non-uniform temperature field \cite{morgan}.

In this paper, we investigate the impact of the electrothermal effect on the DEP-separation of SWNTs by formulating a dynamical model of the integrated system. Here we mainly discuss the transport of s-SWMT, to which electrothermal force has non-trivial effects. We demonstrate that for a commonly used surfactant and electric field with practical magnitude, electrothermal flows can be sufficiently large to have a substantial impact on the separation efficiency of the DEP method. It is shown that electrothermal flows can significantly weaken the DEP-separation by driving the s-SWNTs towards the electrodes, which is consistent with the experimental observations.
\section{Mathematical modeling}
%%%%%%%%%  Simulation methods  %%%%%%%%%%%
Using data for a commonly used material in experiments \cite{krupke2004}, the system under consideration consists of SWNTs with a diameter of 1.4 nm and a length of 1 $\mu$m dispersed in aqueous solutions of sodium dodecylbenzene sulfonate (SDBS). For the electrical conductivity, we adopt the effective conductivity (0.35 S/m) suggested by Krupke {\it et al} \cite{krupke2004}. While the influence of surfactant on the electrical conduction of SWNTs surrounded by SDBS is not clear, we adopt the empirically effective value. Similarly, the permittivity of s-SWNTs was set to be $5\epsilon_0$ \cite{krupke2004}. Although the magnitude of permittivity of s-SWNTs is arguable, a variation within one order of magnitude should not affect the current analysis.

Electrical conductivity of SDBS solutions $\sigma_\text{m}$ was estimated as 4 mS/m, 29 mS/m and 230 mS/m for concentration of $0.01\%$, $0.1\%$ and $1\%$, respectively \cite{krupke2004}. The permittivity of the SDBS solutions $\epsilon_\text{m}$ at room temperature is about $80\epsilon_0$ and has negligible dependence on the concentration. It is straightforward to calculate the DEP force factor for s-SWNTs and the results show that it decreases about one order when the concentration of SDBS increases by one order.

\begin{figure}[pt]
\subfigure[Geometry of electrodes used in calculations]{
\includegraphics[width=4.cm]{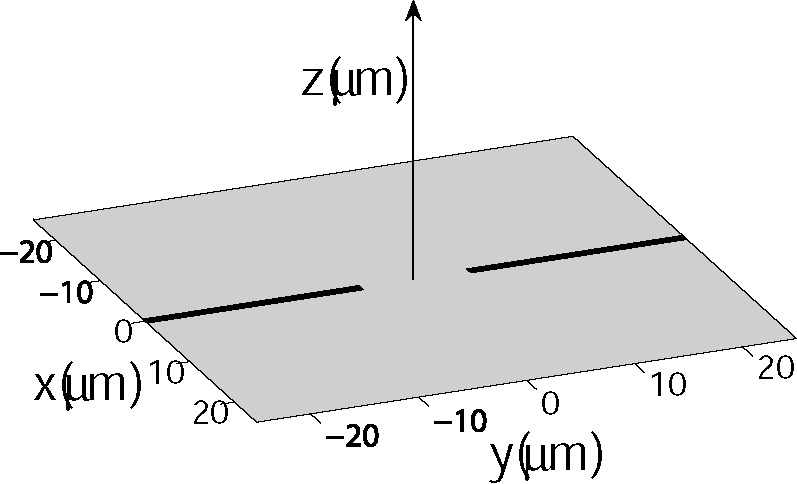}
\label{fig1a}}
\subfigure[Temperature changes with $z$ at ($x,y$)=($0~\mu$m$,5~\mu$m), $\phi$ is 20 V, $\omega$=300 kHz]{
\includegraphics[width=4.1cm]{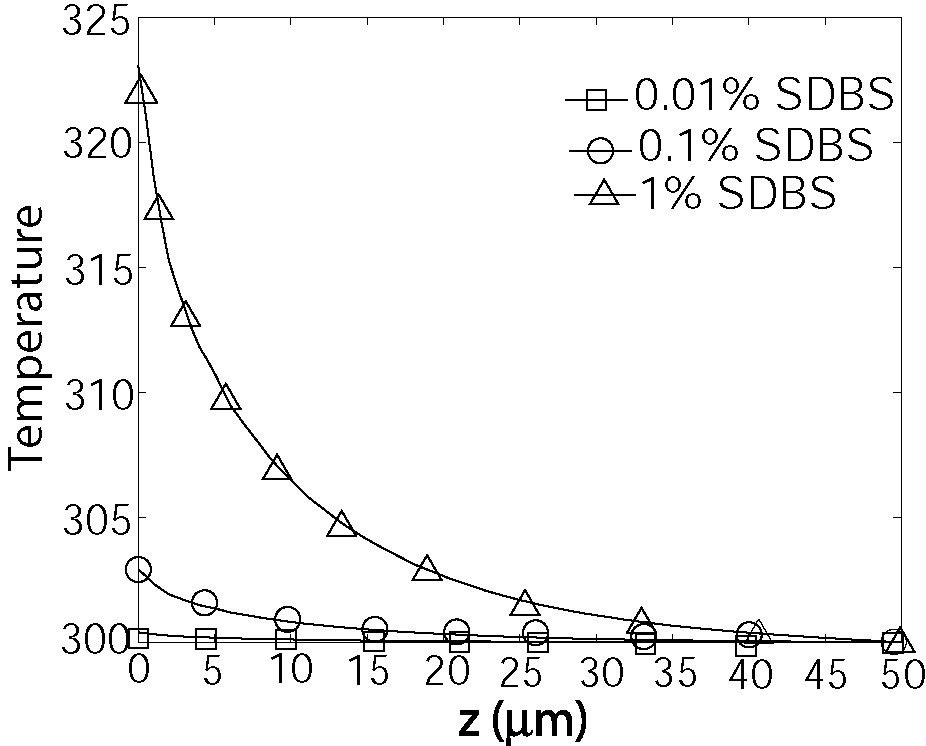}
\label{fig1b}} \caption{Geometry and temperature profiles.}
\end{figure}

We solve the heat conduction equation with a heat source due to an electric field \cite{ramos}, $k\nabla^2T+\sigma_\text{m} E^2=0$, where the thermal conductivity $k$=0.6 Wm$^{-1}$K$^{-1}$. As a typical system geometry in experiments \cite{krupke2004}, a pair of 20 $\mu$m-long and 1 $\mu$m-wide electrodes were located with a distance of 10 $\mu$m at the bottom of the calculation domain as shown in Fig.~\ref{fig1a}.
The boundary conditions are: $T_{\text{wall}}=300$ K at the surrounding walls, and $\displaystyle\frac{dT}{dz}=0$ at the bottom, which means the substrate and electrodes are thermally insulated. Figure~\ref{fig1b} shows the temperature profiles for various SDBS concentrations above the tip of the electrode ($x,y$)=($0~\mu$m$,5~\mu$m), with the applied AC potential $\phi=20$ V (peak to peak). Maximum temperature differences are 0.3 K, 1.6 K and 11.8 K on the electrode (z=0) for 0.01\%, 0.1\% and 1\% SDBS solutions, respectively.

% figure 1 is here!

The electrothermal body force acting on the bulk fluid due to the AC electric field is expressed as \cite{ramos}
\begin{equation}
\mathbf
f=-\frac{1}{2}[(\frac{\nabla\sigma_\text{m}}{\sigma_\text{m}}-\frac{\nabla\epsilon_\text{m}}{\epsilon_\text{m}})\cdot\mathbf
E\frac{\epsilon_\text{m}\mathbf E}{1+(\omega\tau)^2}+\frac{1}{2}|\mathbf
E|^2\nabla\epsilon_\text{m}], \label{eqn:3}
\end{equation}
where $\tau=\epsilon_\text{m}/\sigma_\text{m}$ is the charge relaxation time of the solution. The first term on the right hand side expresses the Coulomb force and the second the dielectric force. For a typical electrolyte solution, $(1/\sigma_\text{m})/(\partial \sigma_\text{m}/\partial T)=0.02$~K$^{-1}$ and $(1/\epsilon_\text{m})/(\partial \epsilon_\text{m}/\partial T)=-0.004$~K$^{-1}$ \cite{morgan}. The first (second) term is dominant in the low (high) frequency regime (the crossover frequency is $\omega_c=1/\tau$, which is 5.6 MHz and 41 MHz for 0.01\% and 0.1\% SDBS concentrations). In the calculation, we adopted 300 kHz and 1 GHz, which represent the low and high frequency regime, respectively. For low Reynolds numbers, the transport term of the Navier-Stokes (N-S) equations can be neglected.
Hence, the N-S equations become
\begin{equation}
\rho_m\frac{\partial \mathbf u_\text{f}}{\partial
t}=\eta\nabla^2\mathbf u_\text{f}-\nabla p+\mathbf f \qquad \nabla\cdot\mathbf u_\text{f}=0. \label{eqn:4}
\end{equation}

For low Reynolds numbers, the inertial effect can be neglected \cite{morgan}. The terminal velocity of s-SWNTs is calculated as $\mathbf u=\mathbf u_\text{d}+\mathbf u_\text{f}+\mathbf u_\text{b}$, here $\mathbf u_\text{d}=\mathbf F_{\text{DEP}}/f_\text{t}$ with the translational friction coefficient for a prolate ellipsoid given by $\displaystyle f_\text{t}=\frac{6\pi\eta a}{ln(2a/b)}$ \cite{morgan}. Characteristic variables of electrothermal flow are $U^*=\displaystyle\frac{\partial \sigma_{\textrm m}}{\partial T}\cdot\frac{25\epsilon_\text{m}\phi^4}{k(1+(\omega\tau)^2)\eta
L}$, $T^*=\displaystyle\frac{\sigma_\text{m}\phi^2}{k}$, $L$=1 $\mu$m\ and $t^*=\displaystyle\frac{\partial T}{\partial \sigma_{\textrm m}}\cdot\frac{k(1+(\omega\tau)^2)\eta
}{25\epsilon_\text{m}\phi^4}$. Here $\phi$ is the applied peak-to-peak AC voltage.
From the Einstein relation, the Brownian velocity can be derived as
$\mathbf u_b=\sqrt{6D/\ dt}$ with the diffusion coefficient
$D=(k_BT/f_\text{t})$, and $dt$ is the time interval of observation. We define dimensionless variables $\mathbf{\tilde u}=\mathbf
u/U^*$, $\tilde T=(T-T_{\text{wall}})/T^*$, $\tilde
x=x/L$, $\tilde{dt}=dt/\ t^*$ and $\mathbf{\tilde E}=\mathbf EL/\phi$,
dimensionless parameters $ P_1=\displaystyle\frac{(1+(\omega\tau)^2)}{12}$, $
P_2=\displaystyle0.0463\frac{\partial T}{\displaystyle\partial \sigma_{\textrm m}}\frac{ b^2 \alpha ln(\frac{ a}{
b})k(1+(\omega\tau)^2)}{L^2\phi^2}$ and $
P_3=\displaystyle\sqrt{\frac{\partial T}{\partial \sigma_{\textrm m}}\frac{k_B\sigma_\text{m} ln(\frac{2a}{
b})\cdot(1+(\omega\tau)^2)}{0.6\cdot\epsilon_\text{m}\phi^2\pi a}}$.
Consequently, we obtain a set of system equations in a
non-dimensional form,
\begin{eqnarray}
\nabla\cdot\mathbf{\tilde E}&=&0, \quad \nabla^2\tilde T=-|\mathbf{\tilde
E^2}|,\nonumber\\
\frac{\partial \mathbf{\tilde u_\text{f}}}{\partial\tilde
t}&=&\nabla^2\mathbf{\tilde u_\text{f}}-\nabla\tilde p+(\nabla\tilde T\cdot
\mathbf{\tilde E}\mathbf){\tilde E}+P_1|\mathbf{\tilde
E^2}|\nabla\tilde T, \nonumber \\
\mathbf{\tilde u_{b}}&=&\tilde r\sqrt{
(\tilde T+T_{\text{wall}}/\ T^*)/\ \tilde{dt}},\quad
\mathbf{\tilde u_\text{d}}=\nabla|\mathbf{\tilde E}|^2 \nonumber \\ \mathbf{\tilde u}&=&P_2\mathbf{\tilde u_\text{d}}+\mathbf{\tilde
u_\text{f}}+P_3\mathbf{\tilde u_\text{b}}, \label{eqn:6}
\end{eqnarray}
where $\tilde r$ denotes a random number with normal distribution whose mean is 0 and standard deviation is 1. Equation (4) is solved using finite element method toolbox femLego\cite{gustav}.

\section{Force estimations}
\begin{figure}[tbp]
\centering
\includegraphics[width=8.2cm]{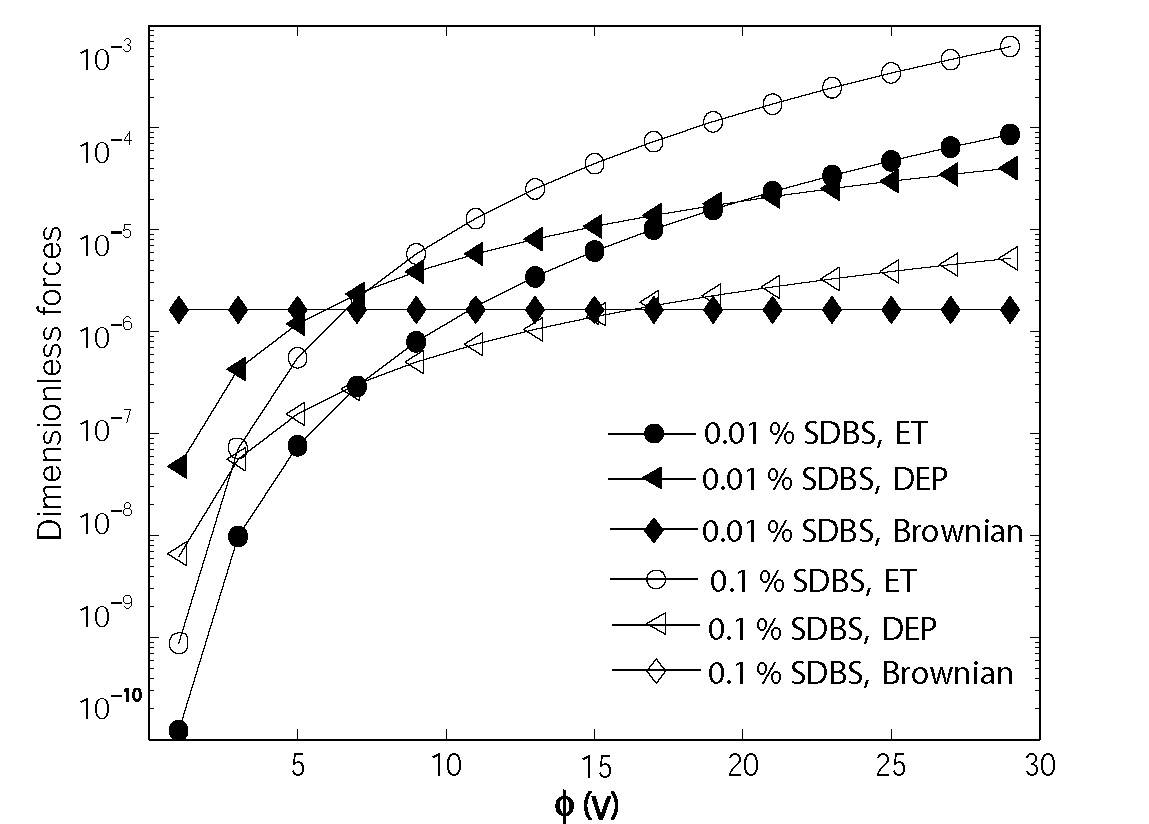}
\caption{Forces change with $\phi$ at ($x,y,z$)=($0~\mu$m, $14~\mu$m, $1~\mu$m). The Brownian force profiles overlap on each other}
\label{fig2}
\end{figure}

Since the inertial effect is neglected, the dimensionless velocities are equivalent to the dimensionless forces. Taking the characteristic timescale $t^*$ as the time interval $dt$, the dependence of the relative magnitudes of DEP, Brownian and electrothermal forces on the applied electric potential was examined. Figure~\ref{fig2} shows the dimensionless forces at ($x,y,z$)=(0 $\mu $m, 14 $\mu$m, 1 $\mu$m) for two different surfactant concentrations 0.01\% and 0.1\%. The $y$ position was chosen to be in the regime with the positive electrothermal force that attracts SWNTs towards the electrode and $z$=1 $\mu$m equals to the length of SWNTs. Here, the frequency is 300 kHz, which is within the frequency range commonly used in experiments \cite{krupke2004, zhibin1}. Figure~\ref{fig2} shows a clear dependence of the locally dominant force on the applied voltage and surfactant concentration. The change of the Brownian motion due to voltage and surfactant concentration is subtle and relatively negligible. The voltage-dependence of the strength of the electrothermal force is proportional to $\phi^4$ and that of the DEP force is proportional to $\phi^2$. As a result, for both concentrations 0.01\% and 0.1\%, the electrothermal force overcomes the DEP forces at the high voltage limit since the electrothermal force increases with the voltage much faster than the DEP force.

As for the influence of the surfactant concentration, the electrothermal force increases with the concentration while the DEP force takes the opposite trend. When surfactant concentration is 0.01\%, the dominant force is the Brownian force for low voltage ($\phi<6$ V), the DEP force for intermediate voltage (6 V$<\phi<19$ V) and the electrothermal force for high voltage ($\phi>19$ V). On the other hand, when the concentration is 0.1\%, the local DEP force around the selected location is never dominant for the entire range of $\phi$. This implies that for SDBS concentration higher than 0.1\%, the positive DEP force on s-SWNTs plays minor role and the transport is governed mainly by the electrothermal force. The result suggests that there is a crossover between the electrothermal and DEP force with respect to the SDBS concentration. This means that beyond the crossover, an attempt to enhance the DEP-separation efficiency by increasing the surfactant concentration, {\it i.e.} by reducing the magnitude of the positive DEP on s-SWNTs may result in enhancing the electrothermal force, which reduces the efficiency.

%figure 2 is here!

\begin{figure}[tbp] \centering
\includegraphics[width=8.2cm]{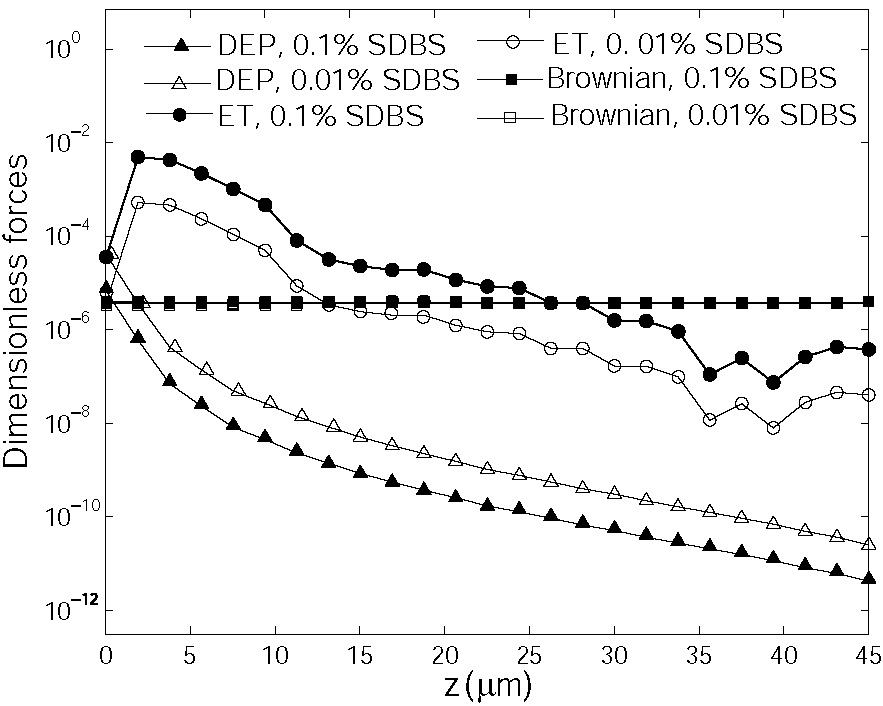}
\caption{Forces change with z, ($x$, $y$)=(0 $\mu$m, 14 $\mu$m), $\phi$=20~V. The Brownian force profiles overlap on each other} \label{fig3}
\end{figure}

Figure~\ref{fig3} shows the magnitudes of the dimensionless forces with the vertical coordinate $z$ for the same ($x$, $y$) locations as in Fig.~\ref{fig2}. Following typical experiments, $\phi$=20 V was applied to 0.1\% and 0.01\% SDBS solutions. Figure~\ref{fig3} shows that in 0.1\% SDBS, the Brownian motion surpasses the DEP force except very close to the electrode (less than 1 $\mu$m). Here we observe that, in a domain larger than that where the DEP force is dominant, s-SWNTs are carried by the electrothermal flow for both SDBS concentrations. As a result, the electrothermal flow overcomes the Brownian motion in a relatively large domain and transport s-SWNTs onto the electrodes. To demonstrate this, we will verify the direction of the electrothermal flow and the actual traces of s-SWNTs under the action of the resultant force.
%figure3

\section{Transport simulations of semiconducting SWNTs}
\begin{figure}[tbp]
 \subfigure[Streamlines of the electrothermal flow, $\omega$=300~kHz]{
 \includegraphics[width=3.9cm]{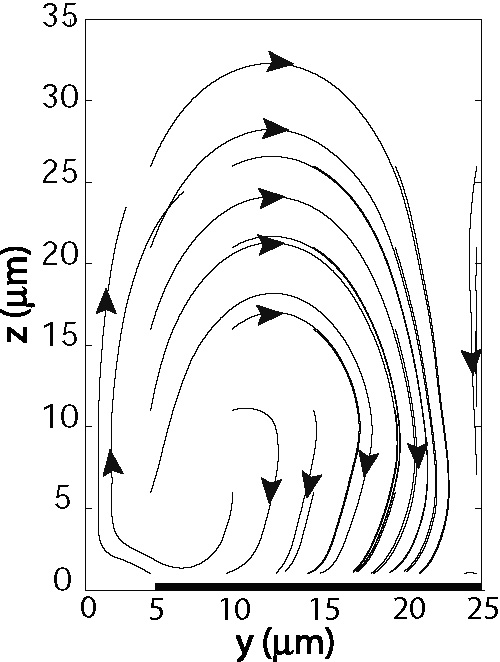}
 \label{fig4a}}
 \subfigure[Streamlines of the electrothermal flow, $\omega$=1~GHz]{
 \includegraphics[width=3.9cm]{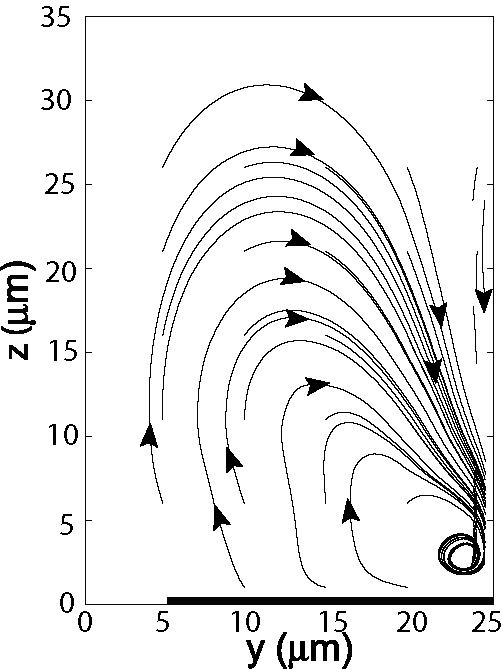}
 \label{fig4b}}\\
 \subfigure[Traces of s-SWNTs, $\omega$=300~kHz]{
 \includegraphics[width=3.9cm]{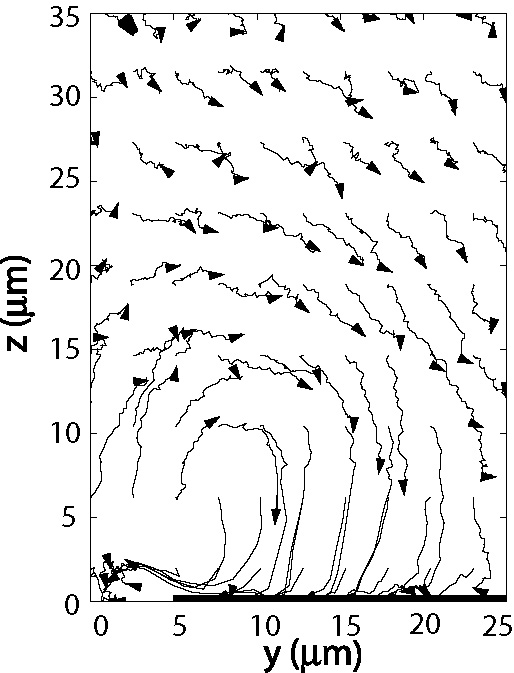}
 \label{fig4c}}
 \subfigure[Traces of s-SWNTs, $\omega$=1~GHz]{
 \includegraphics[width=3.9cm]{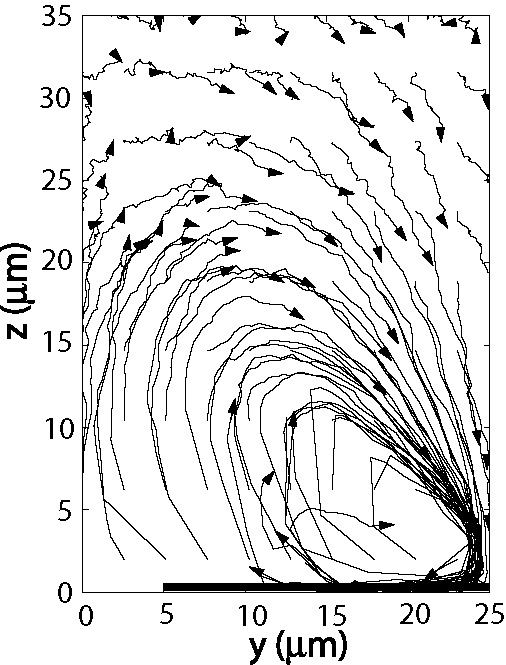}
 \label{fig4d}}
 \caption{Streamlines and corresponding traces of s-SWNTs in 0.1\% SDBS solutions, low (300~kHz) and high (1~GHz) frequency regimes
 with $\phi$=20~V.}
 \label{fig4}
 \end{figure}

Figures~\ref{fig4a} and \ref{fig4b} show streamlines of the electrothermal flows in 0.1\% SDBS solution for two different frequencies. They are cross-sectional views of the 3D simulation results at the plane of $x=0$. The thick line drawn from ($x, y$)=(0 $\mu$m, 5 $\mu$m) to (0 $\mu$m, 25 $\mu$m) marks one of the two electrodes [Fig.~\ref{fig1a}]. The evident dependence of the flow pattern on $\omega$ is highlighted by comparing the flow pattern for $\omega=300$ kHz [Fig.~\ref{fig4a}] corresponding to a typical working frequency with that for $\omega=1 $ GHz [Fig.~\ref{fig4b}] corresponding to the upper limit of frequency explored in experiments. When $\omega=300$ kHz, the vortex core appears close to the gap of the electrodes, whereas when $\omega=1$ GHz, it is located close the domain boundary ($y$=25 $\mu$m). The pattern of the electrothermal flow varies with frequency due to variation in the balance of the Coulomb force and dielectric force. The flow patterns for $\omega=300$ kHz and $\omega=1$ GHz resemble the flow patterns for low and high frequency limits, where the Coulomb force and dielectric force become dominant, respectively. The flow pattern also strongly depends on the geometry of the system. In the current 3D system, independently of frequency, the electrothermal force gives rise to upward flows on the gap of the electrodes and downward flows on domain boundary ($x=0~\mu$m, $y=25~\mu$m). This differs from the 2D case with an infinitesimal gap, where the direction of the flow circulation at low frequency limit is opposite from that at high frequency limit \cite{ramos,morgan}.
%figure 4 is here!
     
After solving Eq.~(\ref{eqn:6}), we obtain the actual trajectories of the s-SWNTs as plotted in Fig.~\ref{fig4c} and Fig.~\ref{fig4d}. The simulation times are 33 ms and 24 ms respectively. The s-SWNTs are seen to follow the streamlines of the electrothermal flow, especially well in the region close to the electrodes. This agrees with the local force comparisons shown in Fig.~\ref{fig3}, where the electrothermal force appeared to be dominant in a broad region close to the electrode. This means that the variation of the flow patterns observed in Fig.~\ref{fig4a} and Fig.~\ref{fig4b} results in different spatial trapping distributions of SWNTs. When $\omega=300$ kHz, s-SWNTs are attracted broadly on the electrode, while when $\omega=1$ GHz, the s-SWNTs are mainly attracted to a relatively small region close to the domain boundary.

Let us briefly discuss the influence of the electrothermal effect on m-SWNTs. We have also performed the simulations for m-SWNTs with $\sigma_{p}=10^8$~S/m and $\epsilon_{p}=-10^4 \epsilon_0$ \cite{dimaki}. The magnitude of the positive DEP force for m-SWNTs is much larger than that for s-SWNTs. Consequently, the impact of both electrothermal and Brownian forces on the overall transport is much smaller for m-SWNTs than for s-SWNTs. However, the electrothermal force is far from negligible, especially for higher SDBS concentrations. In fact, in the gap region of the electrode pair where the electrothermal force is directed upwards ($y$=0 $\mathrm\mu$m to 5 $\mathsf\mu$m in Fig.~\ref{fig4}), the electrothermal force overcomes the DEP force and repels m-SWNTs away from the electrodes when the concentration of SDBS is larger than 0.1\%. The impact on the collection of m-SWNTs may be limited since the m-SWNTs will be circulated along the vortex and eventually transported onto the electrodes after a certain time as seen in Fig.~\ref{fig4} for s-SWNTs. However, the effect should certainly influence the distribution of yielded m-SWNTs on the electrode, which is important for deposition of SWNTs to selected sites for electric device constructions.

\section{Conclusions}
We formulated a dynamical system to simulate the DEP  motion of s-SWNTs, which is different from previous simulations mainly by taking the electrothermal flow into account. To realize the DEP  separation, the applied electric potential needs to be high enough so  that the DEP force on m-SWNTs overcomes the Brownian motion. This  results in a high temperature gradient, which creates a substantial  electrothermal flow. This electrothermal flow will bring the s-SWNTs close to
the electrodes  where the weaker DEP force on s-SWNTs  can collect them. Thus, the  collaborative action of electrothermal flow and DEP restricts the allowable range  of potentials where DEP separation may be possible. The higher the  concentration of the surfactant is, the stronger the electrothermal flow is.  Also, when the frequency increases, the pattern (direction and  magnitude) of the electrothermal flow changes. Therefore the electrothermal flow would very likely increase the difficulty both in separating s-SWNTs from m-SWNTs and in depositing SWNTs to a certain position. The main  conclusion of this paper is that when designing DEP separation of  SWNTs, it is necessary to consider the possible influence of the electrothermal flow.

The Swedish Research Council (VR) is greatly acknowledged for financial support. The authors thank Z.-B. Zhang and S.-L. Zhang for stimulating discussions.

\bibliography{DEPbib}

\end{document}